# A Mathematical Trust Algebra for International Nation Relations Computation and Evaluation

**Mohd Anuar Mat Isa** (*corresponding author[1])
**Ramlan Mahmod**
**Nur Izura Udzir**
Faculty of Computer Science and Information Technology, Universiti Putra Malaysia, Malaysia

**Jamalul-lail Ab Manan**, Section 7, Shah Alam, Malaysia.
**Ali Dehghan Tanha**, University of Salford, The Crescent, Salford, United Kingdom.

**Abstract**

This manuscript presents a trust computation for international relations and its calculus, which related to Bayesian inference, Dempster-Shafer theory and subjective logic. We proposed a method that allows a trust computation which is previously subjective and incomputable. An example of case study for the trust computation is the United States of America–Great Britain relations. The method supports decision makers in a government such as foreign ministry, defense ministry, presidential or prime minister office. The Department of Defense (DoD) may use our method to determine a nation that can be known as a friendly, neutral or hostile nation.

## Introduction

This publication describes an extension of our previous works related to trust issues in international nation relations. In our previous works (Mohd Anuar Mat Isa et al., 2012a, 2012b), we have mentioned the need for a "*trust model*" in Common Criteria (CC). In this work, we model the international relations between nations using a calculus model, which we call a trust algebra. The proposed method will allow a trust computation, which is previously subjective and incomputable.

## Related Work

There are many trust definitions that were expressed in natural languages. To enable trust computation in computing systems, the trust definitions need to be conveyed in a measurable or quantifiable notations such as statistical representations. The first attempt to represent trust in the statistical notations was by (Dempster, 1967). Dempster showed a probability measurement that is to define an upper and lower probabilities for a multivalued mapping. The probability measurement is a generalization of calculus in Bayesian theory. His statistical scheme is adopted by (Shafer, 1979, 1976) and it provides elegant method to compute trust. Many researchers later (in 1980-1995) addressed both works as the foundation for a concrete trust computation, which they began to call as Dempster-Shafer theory in the early 1980s (Shafer, 2015). Later (Jøsang, 1997) provided an extension to a probabilistic calculus for the Dempster-Shaper theory by introducing an artificial reasoning, named subjective logic. The following subsections will further discuss the Bayesian theory, Dempster-Shafer theory and subjective logic.

---

[1] This manuscript is a draft version. The final version will be published in a reputable journal. One may contact the main author (anuarls@hotmail.com) for further clarification or discussion for an implementation of the trust algebra.

## Trust Algebra: Definition and Notation

**Definition 1**. A nation state is a sovereign nation and recognized by the United Nation (UN). Referring to the UN's Charter ("Charter of the United Nations," 1945):

Chapter I, Articles 1: "*To maintain international peace and security…*" and "*to develop friendly relations among nations based on respect for the principle of equal rights…*". Articles 2: "*…principle of the sovereign equality of all its Members.*".

Chapter II, Articles 4: "*Membership in the United Nations is open to all other peace-loving states which accept the obligations contained in the present Charter…*".

Referring to the UN's Charter, we define a **nation** term in this work as the nation state or any UN member states.

**Definition 2.** Trust relation is a relationship between Nation A and Nation B. The trust relation can be either friendly (ally), neutral, or hostile (enemy). The trust relation $\mathcal{R}_{A,B}$ denotes a *trust perception* of Nation A toward Nation B. Assume that:

$f \stackrel{\text{def}}{=} friendly,\ n \stackrel{\text{def}}{=} neutral, h \stackrel{\text{def}}{=} hostile$
$f, n, h \in RELATION$
$A, B \in NATION$
$\mathcal{R}_{A,B} = (f \cap n \cap h) = \emptyset$
$\mathcal{R}_{A,B} = (f \cap n) \cup (f \cap h) \cup (n \cap h) = \emptyset;$

**Remark 2.1** Trust relation for $\mathcal{R}_{A,A}$ is reflexive with always friendly.
**Remark 2.2** Trust relation for $\mathcal{R}_{A,B} \neq \mathcal{R}_{B,A}$ is not always symmetric.
**Remark 2.3** Trust relation for $\mathcal{R}_{A,B}$ and $\mathcal{R}_{B,C}$ does not always imply that $\mathcal{R}_{A,C}$ is transitive for relations between Nation A, Nation B and Nation C.
**Remark 2.4** Trust relation for $\mathcal{R}_{A,B}$ and $\mathcal{R}_{B,A}$ are commutative for binary operation $(\mathcal{R}_{A,B}, \mathcal{R}_{B,A}) = (\mathcal{R}_{B,A}, \mathcal{R}_{A,B})$ for addition and multiplication operations.

**Definition 3**. Trust relation for Nation A and Nation B is undefined for $\mathcal{R}_{A,B} = (f \cup n \cup h) = \emptyset.$

**Remark 3.1** Trust relation for $\mathcal{R}_{A,B}$ is undefined when a relation between Nation A and Nation B is neither friendly, neutral, nor hostile. The state of the relation is unknown.
**Remark 3.2** If a definition of a *nation* is reduced to Definition 1, the trust relation always exists because of diplomatic relations and recognitions.

**Definition 4.** Weightage is used for a *linear normalization of trust perceptions* between Nation A toward Nation B. The weightage will help to identify the significance of each trust perceptions.

**Theorem 1.** Mass Weightage
Assume that:
$x \in \mathbb{Z}, \quad x \geq 1;$
$\mathcal{C} = |RELATION|\ or\ cardinality(RELATION)$
Mapped matrix: $\boldsymbol{RELATION} \times (\mathbf{1} \leq \boldsymbol{X} \leq \boldsymbol{C})^T$

$[\boldsymbol{f} \quad \boldsymbol{n} \quad \boldsymbol{h}] \times \begin{bmatrix} 1 \\ 2 \\ 3 \end{bmatrix}\ s.t. (f \mapsto 1), (n \mapsto 2)\ and\ (h \mapsto 3).$

$\mathcal{W}_{Mass} = \sum_{x \leq \mathcal{C}}^{x=1} \mathcal{W}_x\ = 1, \quad \mathcal{W}_x \in \mathbb{R}, \quad 0 \leq \mathcal{W}_x \leq 1$

$e.g. when\ x = 1\ ,then\ \mathcal{W}_x\ is\ for\ friendly's\ weightage$

**Remark 4.1** One may choose to use a priori probability to evaluate (assign value) for every $\mathcal{W}_x$. Given that a cardinality is equal to three, then each $\mathcal{W}_x$ is equal to $\frac{1}{3}$. One may also to use a different value of $\mathcal{W}_x$ that is based on the number of trust properties $\mathcal{P}$ as mentioned in Theorem 3. For large numbers of the properties $\mathcal{P}$ for a given $\mathcal{W}_x$, the $\mathcal{W}_x$ should be increased to represent large samples of the properties $\mathcal{P}$. However, the value of $\mathcal{W}_x$ is subjective to an observer.

**Definition 5.** Scalar is used to determine an interval scale for international nation relations that are either friendly, neutral, or hostile.

**Theorem 2.** Mass Scalar
Assume that:
$\mathcal{S}_1 = h's\ scalar\ sign = -1$
$\mathcal{S}_2 = n's\ scalar\ sign = +1$
$\mathcal{S}_3 = f's\ scalar\ sign = +1$
 * One may choose scalar signs: either +ve or –ve)[2].
$$\mathcal{S}_{Mass} = \sum_{x \leq \mathcal{C}}^{x=1} |\mathcal{S}_x . \mathcal{W}_x| = 1$$

**Lemma 2.1.** Lower bound, middle bound and upper bound in Mass Scalar (interval scale).
$\mathcal{S}_{Lower} = \mathcal{S}_1 . \mathcal{W}_1$
$$\mathcal{S}_{upper} = \sum_{x \leq \mathcal{C}}^{x=2} \mathcal{S}_x . \mathcal{W}_x$$
$\mathcal{S}_{Lower} + \mathcal{W}_1 \leq \mathcal{S}_{middle} \leq (\mathcal{S}_{upper} - \mathcal{S}_\mathcal{C} . \mathcal{W}_\mathcal{C})$

**Definition 6.** Trust perception is a collection of trust properties or elements that are used in determining a trust alignment for $\mathcal{R}_{A,B}$.

**Remark 6.1** Trust relations $\mathcal{R}_{A,B}$ will have the collection of trust properties $\mathcal{P}$ for each trust perceptions (e.g. $f_\mathcal{P}, n_p, h_p$). Each trust property $\mathcal{P}_x$ can be mapped into nominal data with values such as military 0.2, politic 0.3, trade 0.1, spying 0.05, etc.

**Theorem 3.** Mass Trust Properties
 Assume that:
$\mathcal{P}_x \in \mathbb{R}, \quad s.t.\ 0 \leq \mathcal{P}_x \leq 1;$
Let cardinalities: $i, j, k$
$i = |friendly\ properties|$
$j = |neutral\ properties|$
$k = |hostile\ properties|$
$$h_{\mathcal{P}_{Mass}} = \sum_{x=k}^{x=1} \mathcal{P}_x \leq 1$$

---

[2] We choose to use a negative sign for a hostile and positive sign for a neutral and friendly relations. In common sense, the negative sign may suitable to be used for the hostile relation.

$$n_{\mathcal{P}_{Mass}} = \sum_{x=j}^{x=1} \mathcal{P}_x \leq 1$$

$$f_{\mathcal{P}_{Mass}} = \sum_{x=i}^{x=1} \mathcal{P}_x \leq 1$$

**Definition 7.** Trust relations for $\mathcal{R}_{A,B}$ is a product of Mass Trust Perception $\mathcal{T}_{Mass}$. The Mass Trust Perception is a point that resides in a relative distance between a lower bound and upper bound of Mass Scalar. To determine the trust relations for the $\mathcal{R}_{A,B}$, i.e. either friendly, neutral or hostile:
- If the point falls into less than middle bound, it is a hostile relation;
- If the point falls into greater than middle bound, it is a friendly relation;
- If the point falls into the middle bound, it is a neutral relation.

**Remark 7.1** Theorems 1 through 4 rely on three major conditions of relations between nations (hostile, neutral and friendly). One may define more than triple conditions to implement granularity and fuzziness in the relations.

**Remark 7.2** One should not modify in order to implement an additional condition of relations because it will increase difficulties in properties $\mathcal{P}$ classification and nominal data (value assignment). Our suggestion for more than the triple scalar relations is to directly map the Mass Trust Perception's value in Theorem 4 into the septuple scalar. One must define a lower bound and an upper bound for each new relation element. The new relation element is a subset of the existing triple (e.g. Near-Hostile $\subset$ Hostile).

**Theorem 4.** Mass Trust Perception

Mapped matrix ***Mass Trust Properties*** $\times$ **Scalar**

$$\mathcal{T}_{Mass} = \sum \left( [f_{\mathcal{P}_{Mass}} \quad n_{\mathcal{P}_{Mass}} \quad h_{\mathcal{P}_{Mass}}] \times \begin{bmatrix} \mathcal{S}_1.\mathcal{W}_1 \\ \mathcal{S}_2.\mathcal{W}_2 \\ \mathcal{S}_3.\mathcal{W}_3 \end{bmatrix} \right)$$

$$\mathcal{R}_{A,B}(\mathcal{T}_{Mass}) = \begin{cases} hostile, & \mathcal{T}_{Mass} < \mathcal{S}_{Lower} - \mathcal{W}_1 \\ neutral, & \mathcal{T}_{Mass} = \mathcal{S}_{middle} \\ friendly, & \mathcal{T}_{Mass} > \mathcal{S}_{upper} - \mathcal{S}_\mathcal{C}.\mathcal{W}_\mathcal{C} \end{cases}$$

**Lemma 4.1.** Strength of Mass Trust Perception

$$\mathcal{T}_{Mass\,Strength} = \sum \left( [f_{\mathcal{P}_{Mass}} \quad n_{\mathcal{P}_{Mass}} \quad h_{\mathcal{P}_{Mass}}] \times \begin{bmatrix} \mathcal{W}_1 \\ \mathcal{W}_2 \\ \mathcal{W}_3 \end{bmatrix} \right)$$

When $\mathcal{T}_{Mass\,Strength}$ is near to 1, $\mathcal{T}_{Mass}$ may represent many contradiction of opinions between hostile and friendly relations. This may happen if it involves a long duration of sampling (or observation) of international relations between two nations. If the $\mathcal{T}_{Mass}$ is a product of 20 years observation of the international relations between two nations, it may consist of a year of war, a year of military allies, a year of politics disagreement, a year of economy sanctions, etc. If the $\mathcal{T}_{Mass}$ is a product of shorter years observation, the contradiction of opinions may occur when a nation leader or ruling party was changed due to election, revolution, installation of puppet leader as a post-war outcome, etc.

When $\mathcal{T}_{Mass\,Strength}$ is near to 0.5 (or middle), $\mathcal{T}_{Mass}$ may represent a fair opinion that either hostile or friendly relations. If the $\mathcal{T}_{Mass}$ is a product of observation for many years, it may represent consistent international relations during that duration. When $\mathcal{T}_{Mass\,Strength}$ is near to $n_{\mathcal{P}_{Mass}}$, $\mathcal{T}_{Mass}$ represents a bias to a neutral. If the $\mathcal{T}_{Mass}$ is a product of observation for many years, it may represent a firm of neutral relations at that moment. When $\mathcal{T}_{Mass\,Strength}$ and $n_{\mathcal{P}_{Mass}}$ are identical in a positive value, it indicates that there is no hostile property in the calculation (or observation).

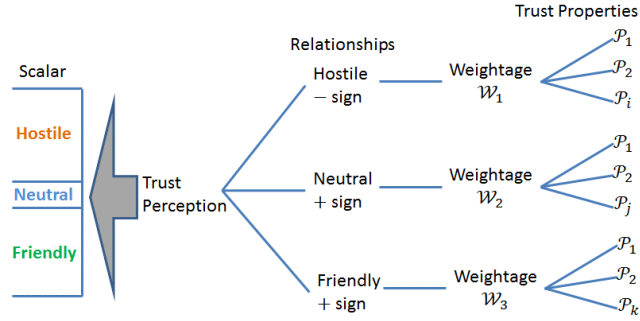

**Fig. 1.** Summary Trust Algebra for Relation Computation

Figure 1 shows a summary of trust computation between two nations. To initialize the trust computation, one should identify trust properties $\mathcal{P}$ and mass weightage for each relation (e.g. Table 1). Let's observe the following example:

Table 1: Trust Computation Example

| Relation | Hostile | Neutral | Friendly |
|---|---|---|---|
| Properties $\mathcal{P}_x$ | $\mathcal{P}_1 = 0.5$<br>$\mathcal{P}_2 = 0.3$<br>$\mathcal{P}_3 = 0$<br>$\mathcal{P}_4 = 0.15$<br>$\mathcal{P}_5 = 0.05$<br>$\mathcal{P}_6 = 0$ | $\mathcal{P}_1 = 0.5$<br>$\mathcal{P}_2 = 0$<br>$\mathcal{P}_3 = 0$<br>$\mathcal{P}_4 = 0.1$ | $\mathcal{P}_1 = 0.05$<br>$\mathcal{P}_2 = 0$<br>$\mathcal{P}_3 = 0$<br>$\mathcal{P}_4 = 0$<br>$\mathcal{P}_5 = 0$<br>$\mathcal{P}_6 = 0.1$ |
| Weightage $\mathcal{W}_{Mass}$ | 0.45 | 0.10 | 0.45 |
| Scalar Sign $\mathcal{S}_x$ | $\mathcal{S}_1$ is $-$ | $\mathcal{S}_2$ is $+$ | $\mathcal{S}_3$ is $+$ |

$h_{\mathcal{P}_{Mass}} = 0.9 = 0.5 + 0.2 + 0.15 + 0.05$
$n_{\mathcal{P}_{Mass}} = 0.6 = 0.5 + 0.1$
$f_{\mathcal{P}_{Mass}} = 0.15 = 0.05 + 0.1$
$\mathcal{S}_{Mass} = 1 = 0.45 + 0.10 + 0.45$
$\mathcal{S}_{Lower} = -0.45$
$\mathcal{S}_{upper} = 0.55 = 0.10 + 0.45$
$-0.45 + 0.45 \leq \mathcal{S}_{middle} \leq (0.55 - 0.45)$
$0 \leq \mathcal{S}_{middle} \leq 0.10$

$\mathcal{T}_{Mass} = \sum([0.9 \quad 0.6 \quad 0.15] \times \begin{bmatrix} -.45 \\ +.10 \\ +.45 \end{bmatrix})$

$\mathcal{T}_{Mass} = -0.2775 = \sum([-0.405 \quad +0.06 \quad +0.0675])$

$\mathcal{R}_{A,B}(-0.2775) = \begin{cases} \textbf{hostile}, & \mathcal{T}_{Mass} \textbf{ is} < \textbf{S\_middle} \\ neutral, & \mathcal{T}_{Mass} \text{ is } S\_middle \\ friendly, & \mathcal{T}_{Mass} \text{ is} > S\_middle \end{cases}$

$\mathcal{T}_{Mass\ Stength} = 0.5325 = \sum([0.9 \quad 0.6 \quad 0.15] \times \begin{bmatrix} 0.45 \\ 0.10 \\ 0.45 \end{bmatrix})$

Referring to Theorem 4 and Lemma 4.1, the given example has shown that a relation between Nation A and Nation B is hostile. The strength of the relation is near to 0.5 such that it represents a consistent hostile relation during the observation.

## Case Study: International Relations

In this section, we explore international nation relations between the United States of America and Great Britain (USA–GB.

*Properties*

We have clustered events that may affect international nation relations as showed in Tables 2, 3 and 4. Clustering or grouping the related events for certain properties will reduce complexities for determining properties' values and it will help to reduce the searching time of the whole data in public domains (e.g. Internet, news, etc.). If at least a single event is found to be related to the given properties, then the given properties will be included in a trust computation. It may not be strong enough as a solid evidence for the given properties, but it will help to enable the trust computation. The Dempster-Shafer's theory of evidence may also be applied in event verifications. However, it requires too much effort.

Table 2. Friendly (Positive)

| $\mathcal{P}_x$ | Descriptions |
|---|---|
| $\mathcal{P}_1$ 0.5 | War ally and mutual defense pact during war. |
| $\mathcal{P}_2$ 0.2 | Share/trade nuclear technologies and materials (e.g. uranium) or mass destruction weapon for warfare. Arm collaboration in R&D for warfare. Financial aid for warfare. |
| $\mathcal{P}_3$ 0.1 | Head of the state political sentiment and relationships. |
| $\mathcal{P}_4$ 0.1 | Loan or share strategic technologies and equipment. Civil nuclear trade and agreement. Defense pact that enable during peace. |
| $\mathcal{P}_5$ 0.075 | Share military intelligent. Large scale of joint military drills. |
| $\mathcal{P}_6$ 0.025 | Global War on Terrorism (GWOT) |
| 1.0 | TOTAL |

Table 3. Neutral

| $\mathcal{P}_x$ | Descriptions |
|---|---|
| $\mathcal{P}_1$ 0.25 | Member of UN or nation state recognized by UN. |
| $\mathcal{P}_2$ 0.35 | Economic cooperation. E.g. Bilateral trade, multilateral open market, free trade. |
| $\mathcal{P}_3$ 0.40 | Diplomatic mission (embassy or representative). Disaster aid and peacekeeping. |
| 1.0 | TOTAL |

Table 4. Hostile (Negative)

| $\mathcal{P}_x$ | Descriptions |
|---|---|
| $\mathcal{P}_1$ 0.5 | War Enemy |
| $\mathcal{P}_2$ 0.2 | Strong disapproval of share/trade/usage nuclear technologies and materials, or mass destruction weapon. E.g. nuclear testing, intercontinental ballistic missile (ICBM) development and testing, and arms races. |
| $\mathcal{P}_3$ 0.075 | Economy blockage or sanction. Embargo or boycott. (e.g. large scale product boycott, ban visa) |
| $\mathcal{P}_4$ 0.125 | Closed border military aggressive or hostility. Including land, air, maritime trespassing and terrorism. *peaceful dispute through international law is not included. |
| $\mathcal{P}_5$ 0.05 | Political sentiments and threat by the head of state. |

| $\mathcal{P}_6$ 0.05 | Kill or arrest another nation diplomats. |
| | Espionage. (e.g. spying and hacking) |
| 1.0 | TOTAL |

*Weightage*

We chose to implement 40%:20%:40% as weightages for hostile, neutral and friendly relations. The weightage percentages were decided based on the number of properties for the given relations.

*A Case Study: The USA and GBR (2001-2005)*

The United States of America and Great Britain enjoy a long lasting of good international relations. The British-America (or Anglo-American) relation remains intact as close military allies since the World War II. Both nations also share various strategic information (e.g. UKUSA Agreement (NSA, 2013)).

Table 5. The USA and GBR

| $\mathcal{R}_{USA,GB}$ | Hostile | Neutral | Friendly |
|---|---|---|---|
| Properties $\mathcal{P}_x$ | $\mathcal{P}_1 = 0$ | $\mathcal{P}_1 = 0.25$ | $\mathcal{P}_1 = 0.5$ |
| | $\mathcal{P}_2 = 0$ | $\mathcal{P}_2 = 0.35$ | $\mathcal{P}_2 = 0$ |
| | $\mathcal{P}_3 = 0$ | $\mathcal{P}_3 = 0.40$ | $\mathcal{P}_3 = 0.1$ |
| | $\mathcal{P}_4 = 0$ | | $\mathcal{P}_4 = 0$ |
| | $\mathcal{P}_5 = 0$ | | $\mathcal{P}_5 = 0.075$ |
| | $\mathcal{P}_6 = 0$ | | $\mathcal{P}_6 = 0.025$ |
| Weightage $\mathcal{W}_{Mass}$ | 0.40 | 0.20 | 0.40 |
| Scalar Sign $\mathcal{S}_x$ | $\mathcal{S}_1$ is $-$ | $\mathcal{S}_2$ is $+$ | $\mathcal{S}_3$ is $+$ |

$h_{\mathcal{P}_{Mass}} = 0$
$n_{\mathcal{P}_{Mass}} = 1$
$f_{\mathcal{P}_{Mass}} = 0.70$
$\mathcal{S}_{Mass} = 1$
$\mathcal{S}_{Lower} = -0.4$
$\mathcal{S}_{upper} = 0.6$
$0 \leq \mathcal{S}_{middle} \leq 0.2$
$\mathcal{T}_{Mass} = 0.48$

$$\mathcal{R}_{USA,GB}(0.48) = \begin{cases} hostile, & \mathcal{T}_{Mass} \text{ is} < \mathcal{S}_{middle} \\ neutral, & \mathcal{T}_{Mass} \text{ is } \mathcal{S}_{middle} \\ \textbf{\textit{friendly}}, & \textbf{\textit{T}}_{\textbf{\textit{Mass}}} \textbf{\textit{ is}} > \mathcal{S}_{\textbf{\textit{middle}}} \end{cases}$$

$\mathcal{T}_{Mass\ Stength} = 0.48$

$\mathcal{R}_{USA,GB}$ showed that the $\mathcal{T}_{Mass}$ and $\mathcal{T}_{Mass\ Stength}$ are identical. When both variables are identical in a positive value, it indicates that there is no hostile properties in the observation. The strength of the relations is near to 0.5, which represents a consistent friendly relations in 2001 until 2005.

## Results and Discussion

We have presented case studies for the trust algebra in Trust Algebra section. Both case studies discussed the international nation relations between USA-GB. The properties and weightages are subjective to the observers. In this work, the properties and weightages chosen by the authors are based on public information available in the literatures (refer to the Literature Review section) and the Internet[3]. Based on limited information on the Internet, we have drawn tentative conclusions for $\mathcal{R}_{USA,GB}$ relations in 2001 until 2005. $\mathcal{R}_{USA,GB_{mass}}$ (0.48) showed positive relations, which fall into a friendly threshold. $\mathcal{R}_{USA,GB_{mass\ strength}}$ (0.48) is equal to $\mathcal{R}_{USA,GB_{mass}}$, which implies that there are no hostile properties in the observation. The given tentative conclusions may change due to new evidence and new events that will be known by the observer in future.

---

[3] We do not obtain or use any material that may lead to actions of a cyber-crime, terrorism, spying or any other illegal activities.

## Conclusion

In this work, we have modeled a trust algebra for international nation relations. The purpose of trust algebra method is to allow trust computations and trust modeling. Previously, there is no such a method to perform the trust computations for international relations which are subjective and unquantifiable. We have also presented the international nation relations between USA-GB as a case study to demonstrate the proposed method in a real-world scenario.